\documentclass[fleqn,12pt,twoside,A4]{article}
\usepackage{espcrc1}

\usepackage{graphicx}
\usepackage{epsfig}
\usepackage[figuresright]{rotating}

\newcommand{\AmS}{{\protect\the\textfont2
  A\kern-.1667em\lower.5ex\hbox{M}\kern-.125emS}}

\title{Fundamental Symmetries and Interactions }

\author{Klaus P. Jungmann\address[KVI]{Kernfysisch Versneller Instituut , \\ 
        Zernikelaan 25, 9747 AA Groningen, The Netherlands}}%

\begin{document}

% typeset front matter
\maketitle

\begin{abstract}
In nuclear physics numerous possibilities exist to
investigate fundamental symmetries and interactions.
In particular, the precise measurements of 
properties of fundamental fermions,
searches for new interactions in $\beta$-decays,
and violations of discrete symmeties offer
possibilities to search for physics beyond standard theory.
Precise measurements of fundamental constants
can be carried out. Low energy experiments  
allow to probe New Physics 
at mass scales far beyond the reach of
present accelerators or such planned  for the future and at which predicted
new particles could be produced directly.

\end{abstract}

%\section{Introduction}

\section{Fundamental Forces and Symmetries} 

Symmetries play an important and crucial role in physics. Global symmetries
give rise to conservation laws and local symmetries 
yield forces \cite{Lee_56}. To date we know four fundamental interactions:
(i)    Electromagnetism, 
(ii)     Weak Interactions,
(iii)    Strong Interactions, and 
(iv)   Gravitation.
These four forces are fundamental in the sense that all
observed dynamical processes in physics can be traced back to one or
a combination of them. Together with fundamental symmetries they from a framework on
which all physical descriptions ultimately rest.

The Standard Model (SM) is a remarkable theory which allows that
Electromagnetic, Weak and many aspects of Strong 
Interactions can be described to astounding precision in one single 
coherent picture. It is a major goal in modern physics to find a unified 
quantum field theory which includes all the four known
fundamental forces in physics. On this way, a satisfactory quantum 
description of gravity remains yet to 
be found and is a lively field of actual activity. 

In this article we are concerned with
important implications of the SM and centrally 
with searches for new,  yet unobserved interactions.
Such are suggested by a variety of speculative models
in which extensions to the present standard theory are introduced
in order to explain some of the not well understood and not well 
founded features in the SM.
Among the intriguing questions in modern physics are the hierarchy of the 
fundamental fermion masses and 
the number of fundamental particle generations.
Further, the electro-weak SM has a rather large number of some 
27 free parameters. All of them need to be extracted from experiments.
It is rather unsatisfactory that
the physical origin of the observed breaking of discrete 
symmetries in weak interactions,
e.g. of parity (P), of time reversal (T) and of 
combined charge conjugation and parity (CP), 
remains unrevealed, although the experimental findings can be well
described within the SM.

The speculative models beyond the present standard theory
include such which involve left-right symmetry, 
fundamental fermion compositeness, new particles, leptoquarks, 
supersymmetry, supergravity and many more. Interesting candidates 
for an all encompassing quantum field theory are string or membrane
(M) theories which in their low energy limit may include supersymmetry.

In the field of fundamental interactions there are
two important lines of activities: Firstly, there are searches for physics beyond the SM in order
to base the description of all physical processes on a conceptually 
more satisfying foundation, and, secondly, the application of solid knowledge
in the SM for extracting fundamental quantities and achieving a description of more 
complex physical systems, such as atomic nuclei. 
Both these central goals can be achieved at upgraded present and novel, yet to be built 
facilities. In this connection a high intensity proton driver would serve to
allow novel and more precise measurements in a large number of
actual and urgent issues in this field \cite{NUPECC_2004}.

Here we can only address a few aspects of a rich spectrum of possibilities. 

\section{Fundamental Fermions}

The Standard Model has three generations of fundamental fermions which
fall into two groups, leptons and quarks.
The latter are the building blocks of hadrons and in particular of
baryons, e.g. protons and neutrons, which consist of three quarks each.
Forces are mediated by bosons: 
the photon, the W$^\pm$- and Z$^0$-bosons, and eight gluons.

\subsection{Neutrinos}

The leptons do not take part in strong interactions. 
In the SM there are three charged leptons (e$^-$, $\mu^-$, $\tau^-$)   and
three electrically neutral neutrinos  ($\nu_e$, $\nu_{\mu}$, $\nu_{\tau}$)) 
as well as their respective antiparticles. 
For the neutrinos eigenstates of mass  ($\nu_1$, $\nu_2$, $\nu_3$) 
and flavour are  different and connected through a mixing matrix
analogous to the Cabbibo-Kobayashi-Maskawa mixing in the quark sector (see \ref{CKM}).
The reported evidence for neutrino oscillations strongly indicate finite $\nu$ masses.
Among the recent discoveries are the surprisingly large
mixing angles $\Theta_{12}$ and $\Theta_{23}$ (see \cite{McDonald_2004,Nakamura_2004}).
The mixing angle $\Theta_{13}$, the phases for CP-violation,
the question whether $\nu$'s are Dirac or Majorana particles
and a direct measurement of a neutrino mass
rank among the top issues in neutrino physics.

\subsubsection{Neutrino Oscillations}

The recent developments in the field of neutrino oscillation
research and evidence for such from solar, reactor , atmospheric and accelerator
neutrino experiments are reviewed in \cite{McDonald_2004} and \cite{Nakamura_2004}
in this volume.
\begin{figure}[hbt]
\label{antenna} 
\begin{minipage}{3in} 
   \centering
   \epsfig{figure=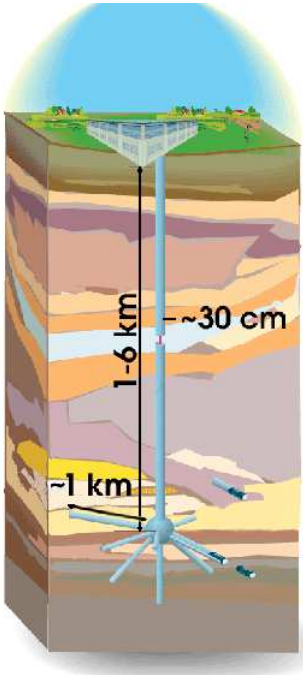,angle=0,width=2.6in}
\end{minipage}       
\hfill
\begin{minipage}{3in} 
   \caption
   {\it A new concept of a direction sensitive 
   neutrino detection for low energy antineutrinos 
   offers not only progress in traditional 
   neutrino research areas like neutrino oscillation, neutrino scattering and 
   supernova watching, but also
   for tomography of the earth to find out about
   the distribution of radionuclides in the earth's crust, mantle and core
   \cite{deMeijer_2004}. The directional sensitivity of this novel
   detector principle  makes it particularly interesting for measuring the
   yet unknown neutrino generation mixing angle $\Theta_{13}$ in a combined 
   near/far detector reactor experiment. }
\end{minipage}
\end{figure}

\subsubsection{Novel Ideas in the Neutrino Field}

Two new and unconventional neutrino detector ideas have come up and gained support in the recent couple of years,
which have a potential to contribute significantly towards solving
major puzzling questions in physics.
\begin{itemize}
\item[(i)] The first concept employs the detection
of high energetic charged particles originating from neutrino reactions 
through Cherenkov radiation in the microwave region (or even sound waves), which results, if 
such particles interact with,e.g., the Antarctic ice or the salt in large salt domes
as they can be found also in the middle of Europe \cite{Gorham_2002}. One advantage of such a detector is its
larger density as compared to water, the typical detector material used up to date. 
It remains to be verified whether this concept will also be applicable for high energetic
accelerator  neutrinos, if timing information and narrowband radio detection techniques will
be employed.
\item[(ii)] The second concept allows directional sensitivity for low energy anti-neutrinos.
The reaction $\overline{\nu} + p \rightarrow e^+ + {\rm n}$ has a 1.8 eV threshold. The resulting
neutron (n) carries directional information in its angular distribution after the event. In
typical organic material the neutron has a range r$_{\rm n}$ 
of a few cm. With a detector consisting of tubes with a
diameter of order r$_{\rm n}$ and with, e.g., boronated walls the resulting $\alpha$-particle from the n+B
nuclear reaction can be used to determine on average the direction of incoming anti-neutrinos.
Such a detector (Fig. \ref{antenna}), if scaled to sufficient mass, can be used to determine the
distribution of radionuclides in the interior of the earth (including testing rather exotic
ideas like the existence of a nuclear reactor in the earth's core). A further 
rather promising application
would be a measurement of the neutrino generation mixing angle $\Theta_{13}$ in  a reactor
experiment with a near and far detector in  $\approx$ few 100~m and  $\approx$ few 100 km 
distance. For this measurement the 
importance of directional sensitivity for low energy $\nu$'s is an indispensable requirement. 
\end{itemize}

\subsubsection{Neutrino Masses}

The best neutrino mass limits result from measurements of the
tritium $\beta$-decay spectrum close to its endpoint.
Since neutrinos are very light particles, a mass measurement can best 
be performed in this region of the spectrum as in other parts the 
nonlinear dependencies caused by the relativistic nature of the kinematic problem 
cause a significant loss of accuracy which overwhelms the gain in statistics
one could hope for. Two groups in Troitzk and Mainz
used spectrometers based on Magnetic  Adiabatic Collimation combined with an Electrostatic filter
(MAC-E technique) and found $m(\nu_e) < 2.2~eV$ \cite{MainzTroitzk,Weinheimer_2003}. 

\begin{figure}[hbt]
\label{KATRIN}
   \centering
   \epsfig{figure=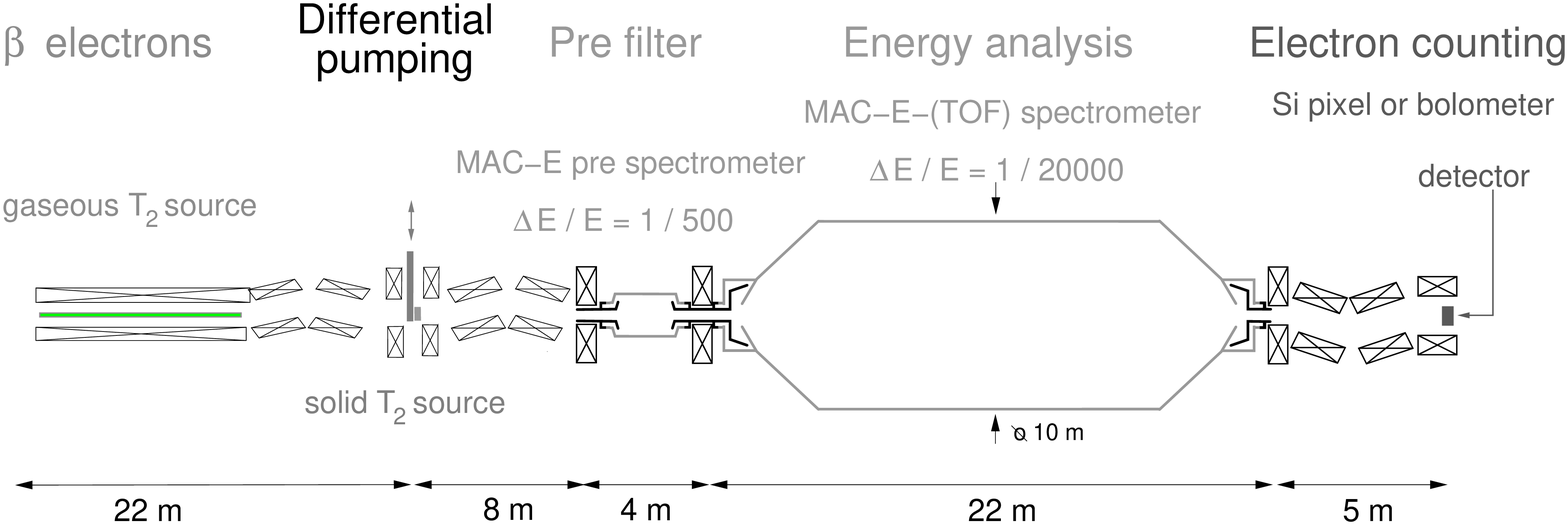,angle=0,width=\textwidth}
   \caption{
   \it
     The KATRIN neutrino experiment aims for measuring a
     neutrino mass directly in a MAC-E spectrometer
     \cite{Weinheimer_2003}. }
\end{figure}

A new experiment, KATRIN \cite{Osipowicz_2001}, is presently prepared
in Karlsruhe, Germany, which is planned to exploit the same technique
(Fig. \ref{KATRIN}). It
aims for an improvement by about one order of magnitude.  The physical dimensions
of a MAC-E device
scale inversely with the possible sensitivity to a finite neutrino mass. 
This  may ultimately limit an approach with this principle. 

The KATRIN experiment will be sensitive to the mass range
where a  finite effective neutrino mass value of between 0.1 and 0.9 eV was
extracted from a signal in neutrinoless double $\beta$-decay in $^{76}$Ge 
\cite{Klapdor_2004}. The Heidelberg-Moskow collaboration performing the
Ge experiment in the Grand Sasso laboratory in Italy reports a 4.2 standard deviation
effect for the existence of this decay \cite{note}. 
It should be noted that neutrinoless double $\beta$-decay is only possible
for Majorana neutrinos. Therefore a confirmed signal would solve one of
the most urgent questions in particle physics.
\begin{figure}[bth]
   \centering
   \epsfig{figure=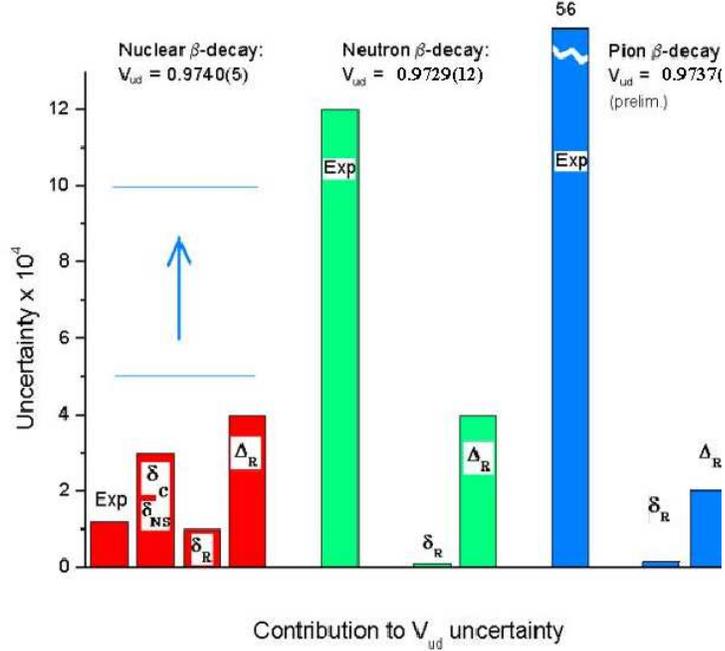,angle=0,width=4.0in}
   \caption
   {\it Uncertainties for three different methods to determine V$_{ud}$: nuclear $\beta$-decays, 
   neutron decays and pion $\beta$-decay. $\delta_R$ is the transition dependent part and 
    $\Delta_R$ is the
    transition independent part of the radiative correction. For
    nuclear $\beta$ -decay there is a radiative correction $\delta_{NS}$ 
    from nuclear structure. The arrow indicates the estimated range of the
    total uncertainty, mainly
    arising from difficulties assigned by the Particle Data Group to calculations of the 
    structure-dependent isospin
    breaking correction  $\delta_C$.} 
\label{vud}    
\end{figure}

\begin{figure}[hbt]
   \centering
   \epsfig{figure=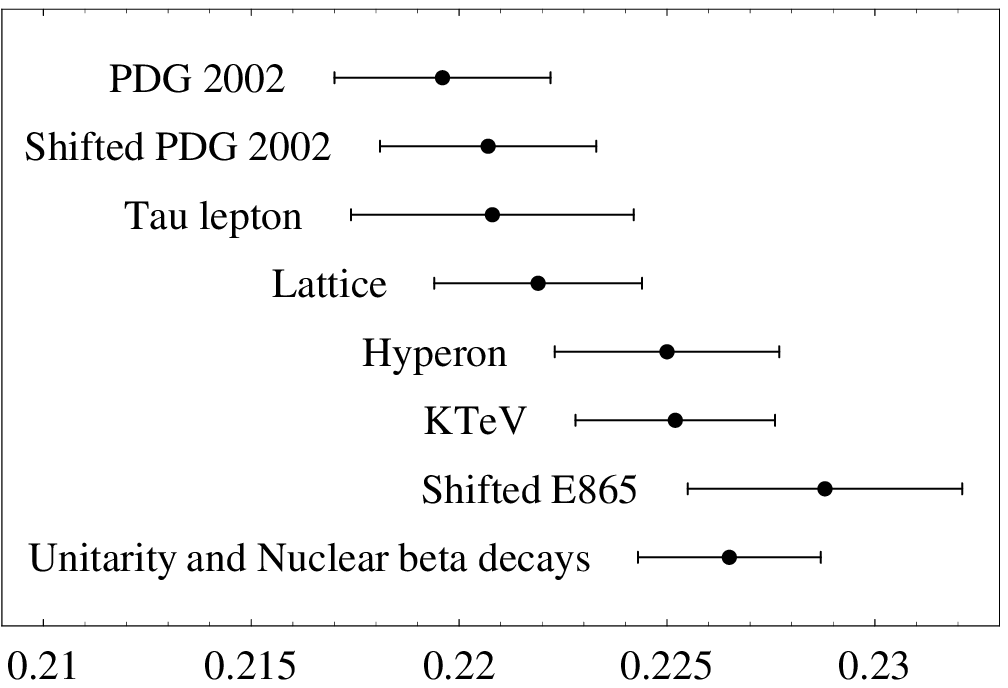,angle=0,width=4.0in}
   \caption[Progress in searches for lepton number violation]
   {\it
     The value of the CKM-matrixelement V$_{us}$ for various  determinations.
     Recent results from several activities in theory and experiment have shown 
     the Particle Data Group
     value for V$_{us}$ to be significantly wrong, due to wrong fit results for 
     K decay branching ratios.
     It appears that together with the precise results from nuclear $\beta$-decays
     the unitarity condition is satisfactorily fulfilled \cite{Czarnecki_2004}.  }
\label{vus}
\end{figure}

\subsection{Quarks - Unitarity of Cabbibo-Kobayashi-Maskawa-Matrix}

\label{CKM}
The mass and weak eigenstates of the six quarks (u,d,s,c,b,t) are different and related to each other
by a $3 \times 3$ unitary matrix, the Cabbibo-Kobayashi-Maskawa (CKM) matrix 
(Table \ref{CKMmatrix}). Non-unitarity
of this matrix would be an indication of physics beyond the SM and could be
caused by a variety of possibilities, including the existence of more than
three quark generations or yet undiscovered muon decay channels. The unitarity of the  CKM matrix
is therefore a severe check on the validity of the standard theory and
sets bounds on speculative extensions to it. 

The best test of unitarity results from the first row of the 
CKM matrix through
\begin{equation}
|{\rm V}_{ud}|^2 + |{\rm V}_{us}|^2 + |{\rm V}_{ub}|^2 = 1 -\Delta ,
\end{equation}
\begin{table}[hbt] 
\caption{\it CKM quark-mixing matrix resulting from recent experiments an their evaluation.}
\label{CKMmatrix} 
\begin{center} 
\begin{tabular}{|l|l|l|}
\hline 
$V_{ud}$ = 0.9735 to 0.9745&  $V_{us}$= 0.2208 to 0.2289&  $V_{ub}$ =  0.0025 to 0.0048\\
\hline 
$V_{cd}$ = 0.219 to 0.226&  $V_{cs}$ = 0.9732 to 0.9748& $V_{cb}$ = 0.038 to 0.044\\
\hline
$V_{td}$ = 0.004 to 0.014& $V_{ts}$ = 0.037 to 0.044& $V_{tb}$ = 0.9990 to 0.9993 \\
\hline
\end{tabular}
\end{center} 
\end{table}
where the SM predicts $\Delta$ to be zero. The size 
of the known elements determine that with the present uncertainties
only the elements V$_{ud}$ and V$_{us}$ play a role. 
V$_{ud}$ can be extracted with best accuracy from the ft values of 
superallowed $\beta$-decays. Other possibilities are the neutron
decay and the pion $\beta$-decay, which both are presently studied (Fig. \ref{vud}).

V$_{us}$ can be extracted from K decays and in principle also from
hyperon decays.
The Particle Data Group \cite{PDG} had decided to increase the uncertainty
   of V$_{ud}$ from nuclear $\beta$-decay \cite{Toner_2003} based on their feelings that
   nuclei would be too complicated objects to trust theory.
   Interestingly, their own evaluation of V$_{us}$ based on Particle Data Group fits
   of K-decay branching ratios turned out to be not in accordance
   with recent independent direct measurements. As a result of the earlier
   too optimistic error estimates in this part a large activity to
   test the unitarity of the CKM matrix took off,
   because a between 2 and 3 standard deviation from unitarity 
   had been persistent. Recent careful analysis of the overall subject 
   has also revealed overlooked inconsistencies in the overall picture
    \cite{Czarnecki_2004,Ellis_2004} and at this time new determinations of
    V$_{us}$ confirm together with V$_{ud}$ from nuclear $\beta$-decay
    that $\Delta=0$ and therewith
    the unitarity of the CKM matrix up to presently possible accuracy (Fig. \ref{vus}).        
   
   Because of the cleanest and therefore
   most accurate theory pion $\beta$-decay (Fig. \ref{vud}) offers for future higher
   precision measurements the best opportunities, in principle.
   The estimate \cite{Hardy_2004} for accuracy improvement from nuclear $\beta$-decays
   is about a factor 2.
   The  main difficulty for new round rests therefore primarily 
   with finding an experimental technique
   to obtain sufficient experimental accuracy for pion $\beta$-decay . 

\subsection{Rare Decays}

In the SM baryon number (B) and lepton number conservation
reflect accidental symmetries.  
A total lepton number (L) and a lepton number for 
the different flavours exists and different conservation laws
were  experimentally established. Some of these 
schemes are additive, some obey multiplicative, i.e.
parity-like, rules.
   
Based on a suggestion by Lee and Yang in 1955 \cite{Lee_56} there is
a strong believe  in modern physics
that a strict conservation of these numbers
remains without a foundation unless they can be
associated with a local gauge invariance and 
with new long-distance interactions
which are excluded by experiments.
Since no symmetry related to lepton numbers could be revealed in the SM,
the observed conservation laws remain without status in physics.
However, the conservation of the quantity (B-L)
is required in the SM for anomaly cancellation. 
Baryon number, lepton number or lepton flavour violation
appear natural in many of the speculative models 
beyond the SM. Often they allow probabilities reaching up
to the present established limits (Table \ref{lnv_limits}).\\

\subsubsection{Lepton Number and Lepton Flavour}

\begin{figure}[thb]
   \centering
   \epsfig{figure=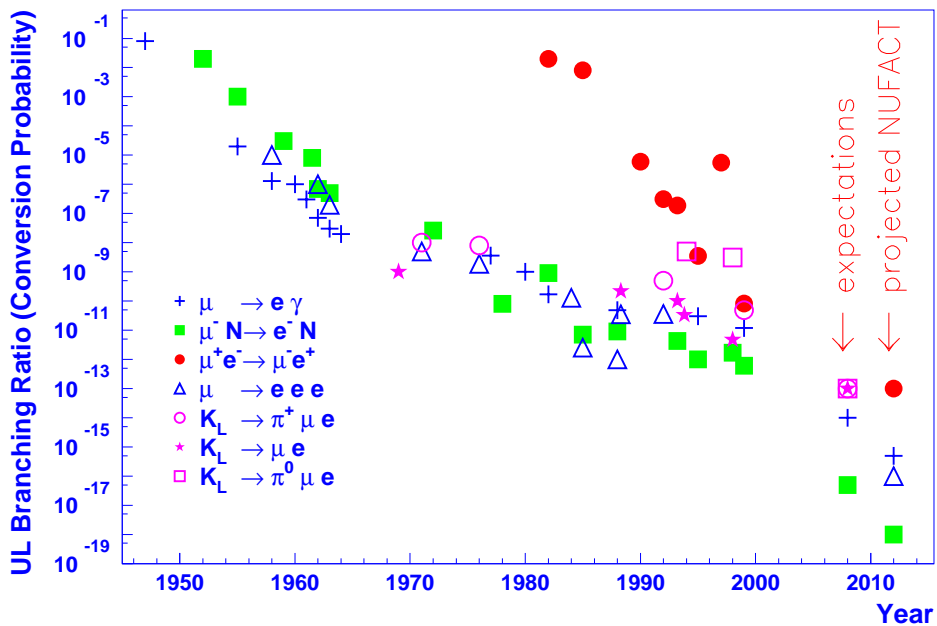,angle=270,width=5 in}
   \caption[Progress in searches for lepton number violation]
   {\protect{\label{lnv_history}}\it
   Dedicated searches for lepton number and lepton flavour  
violating processes
   involving muons ($\mu$) and kaons (${\rm K}$).
   Recent ${\rm K}$ experiments and $\mu^+e^-$ -- $\mu^-e^+$
   conversion show the most significant gain in sensitivity.
   The steady increase in sensitivity is due to both improvements in
   experimental techniques and in the available particle fluxes at 
   accelerators.
   Projections of possibilities of ongoing activities by their 
   experimenters as well as
   those of a CERN working group \cite{Aysto_2001} 
   for a  neutrino factory (NUFACT, 4MW proton driver) are shown.}
\end{figure}

The observations of the neutrino-oscillation experiments
 have demonstrated that
lepton flavour is broken and only the total additive 
lepton number has remained unchallenged.
Searches for charged lepton flavour 
violation are practically not affected in their discovery potential
by these neutrino results. 
For example, in a SM with massive neutrinos the
induced effect of neutrino oscillation into the branching probability 
$P_{\mu \rightarrow e \gamma}$  of the possible decay mode 
$\mu \rightarrow e \gamma$ is of order  \cite{Godzev_1994}
\begin{equation}
P_{\mu \rightarrow e \gamma}=\frac{\Delta m_{\nu_1}^2 -\Delta m_{\nu_2}^2 } {400 eV^2} \cdot 10^{-39}.
\end{equation} 
\begin{table*}[t]\center
 \caption[Limits on Lepton number violating processes]
 {\protect{\label{lnv_limits}}\it Recent upper limits on total
lepton number and lepton flavour violating processes (90\% C.L.). 
Expected limits from ongoing 
experiments and the possibilities at future facilities are given.
 \cite{Aysto_2001}}
{\footnotesize   
\begin{tabular}{|clcccc|}
 \hline
\hspace*{0mm} decay & & actual limit & experiment & present activities & future possibility. \\
 \hline
%Z$^0 $&$\rightarrow \mu e   $ & $1.7  \cdot 10^{-6}$ &LEP&$\approx 10^{-7}$&\\
K$_L$&$\rightarrow \mu e   $ & $4.7  \cdot 10^{-12}$ &BNL E871&&$\approx 10^{-13}$\\
K$_L$&$\rightarrow \pi^0 \mu e $& $3.1  \cdot 10^{-9}$ &KTeV&&$\approx 10^{-13}$\\
K$^+$&$\rightarrow \pi^+ \mu e $& $4.8  \cdot 10^{-11}$ &BNL E865&&$\approx 10^{-13}$\\
$\mu^+$&$\rightarrow e^+ \nu_{\mu} \overline{\nu}_e$ & $2.5  \cdot 10^{-3}$ &KARMEN&&\\
$\mu$&$\rightarrow eee     $ & $1  \cdot 10^{-12}$ &SINDRUM I &&$\approx 10^{-16}$\\
$\mu$&$\rightarrow e \gamma  $   & $1.2 \cdot 10^{-11}$ &MEGA&$2\cdot10^{-14}$&$10^{-15}$\\
$\mu^-$Ti&$\rightarrow e^- $Ti      & $6.1 \cdot 10^{-13}$ &SINDRUM II&
$5\cdot 10^{-17}$ (Al)&$10^{-18}$\\
$\mu^-$Ti&$\rightarrow e^+ $Ca      & $1.7 \cdot 10^{-12}$ &SINDRUM II&&\\
B$^0 $&$\rightarrow  \mu e $ & $5.9 \cdot 10^{-6}$&CLEO&&\\
$\mu^+ e^-$&$\rightarrow \mu^- e^+$ & $ 8.1 \cdot 10^{-11}$&MACS&&$10^{-13}$\\
$\tau$&$\rightarrow e \gamma  $   & $2.7 \cdot 10^{-6} $&CLEO&$\approx 10^{-7}$&\\
$\tau$&$\rightarrow \mu \gamma  $   & $3.0 \cdot 10^{-6} $&CLEO&$\approx 10^{-7}$&\\
$^{76}$Ge &$\rightarrow ^{76}$Se~~ $e^-e^-$ & $0.1 eV < m_{\nu_e}(Maj)$
&HD-MOSCOW& &\\ 
          &                                 &$   <0.9 eV$ &&&\\
\hline
\end{tabular}
}
\end{table*} 
This can be completely neglected in view of present experimental possibilities.
Therefore we have a clean possibility to search for New Physics
at mass scales far beyond the reach of
present accelerators or such planned  for the future and at which predicted
new particles could be produced directly.
The rich spectrum of possibilities is summarized in Table \ref{lnv_limits}
and
the history and future possibilities for lepton flavour and lepton number violating 
processes  is illustrated in Figure \ref{lnv_history}. The future projections
depend strongly on the availability of a new intense source of particles
such as expected from a facility with a high power ($\ge$ MW beam power) proton driver. 

\subsubsection{Baryon Number Violation}

Generally, in most models which aim for the Grand Unification of
all forces in nature baryon number is not conserved. This 
has lead over the past two decades to extensive 
searches for proton decays into various channels.
Present large neutrino experiments have in part emerged form
proton decay searches and such detectors are well suited to
perform these searches along with neutrino detection. 
Up to now numerous decay modes have been investigated 
and partial lifetime limits could
be established 
up to  $10^{33}$ years. 
These efforts will be continued with existing 
setups over the next decade and the detectors 
with the largest mass have highest sensitivity.

An oscillation between the neutron and its antiparticle (n-$\overline{\rm n}$)
would violate baryon number by two units \cite{Chung_2002}. 
Two in principle different approaches have been employed in the 
latest experiments. Firstly, such searches
were performed in the large neutrino detectors, where
an oscillation occurring with neutrons within the nuclei of 
the detector's material could have been
observed as a neutron annihilation signal in which 2~GeV
energy are released in form of pions. Secondly, at ILL 
a beam of free neutrons was utilized. A  suppression of an oscillation
due to the lifting of the energetic degeneracy between 
$n$ and $\overline{n}$ was avoided by a  magnetically well shielded
conversion channel.
Both methods have established a limit of $1.2 \times 10^8$ s 
for the oscillation time.
Significantly improved limits are expected to emerge from
experiments at new 
intense ultra-cold neutron sources.

%\subsection{New interactions in Nuclear and Muon $\beta$-decay}

\section{Discrete Symmetries}

\subsection{Parity}

The observation of neutral currents together with the
observation of parity non-conserva\-tion in atoms were
important to verify the validity of the SM. The fact that 
physics over 10 orders in momentum transfer - from atoms to highest energy scattering -
yields the same electro-weak parameters may be viewed as 
one of the biggest successes in physics to date.
\begin{figure}[t] 
   \centering
   \epsfig{figure=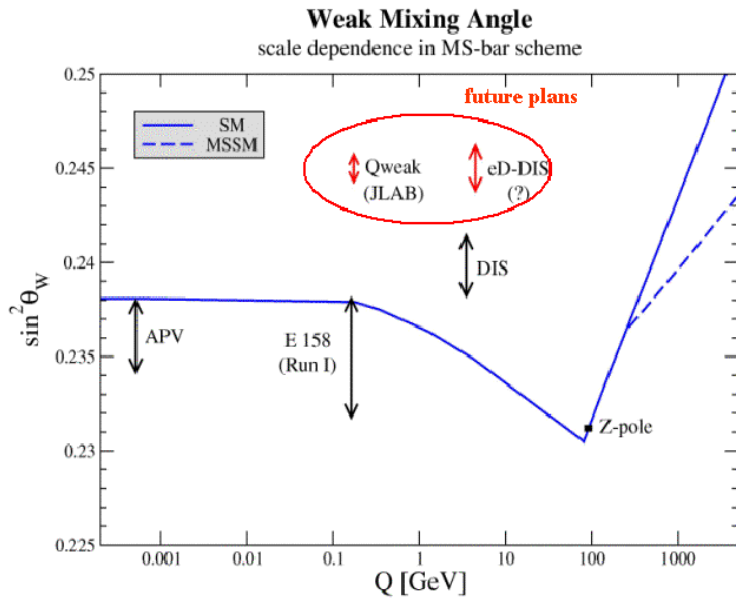,angle=0,width=4in}
   \caption{\it The running of the weak mixing angle $sin \Theta_W$ with 
   energy as predicted by theory \cite{Czarnecki_1998} 
   is not completely reproduced by experiments. Deep inelastic scattering experiments
   appear to show a higher value than predicted. Future planned experiments to clarify 
   the situation are indicated with their expected sensitivity.
   }
\label{stw}  
\end{figure}

However, at the level of highest precision electro-weak experiments
questions arose, which ultimately may call for a refinement.     
The predicted running of the weak mixing angle $sin^2 \Theta_W$ (Fig. \ref{stw})
appears  not to be in agreement with observations \cite{Czarnecki_1998}.
If the value of   $sin^2 \Theta_W$  is fixed at the Z$^0$-pole, deep inelastic electron scattering
at several GeV appears to yield a considerably higher value. A reported disagreement
from atomic parity violation in Cs  has disappeared after a revision of atomic theory.

A new round of experiments is being started with the Q$_{weak}$ experiment \cite{Qweak} at the 
Jefferson Laboratory in the USA.  For atomic parity violation in principle
higher experimental accuracy will be possible from experiments using
Fr isotopes or single Ba or Ra ions in radiofrequency traps \cite{Fortson}.
Although the weak effects are larger in these systems due to their high power dependence on
the nuclear charge, this can only be exploited after better atomic wave function
calculations will be available, as the observation is always through 
an interference of weak with electromagnetic effects.

\subsection{Time Reversal and CP Violation}

The role of a violation of combined charge conjugation (C) and parity (P) 
is of particular importance
through its possible relation to the observed matter-antimatter 
asymmetry in the universe.  This connection 
is one of the strong motivations to
search for yet unknown sources of 
CP violation. A. Sakharov \cite{Sakharov_1967} has suggested that the 
observed dominance of matter
could be explained via CP-violation in the early universe 
in a state of thermal non-equilibrium and with baryon number violating
processes. CP violation as described in the SM 
is insufficient to satisfy the needs of this elegant model.
Permanent Electric Dipole Moments (EDMs) certain
correlation observables  in $\beta$-decays offer excellent opportunities 
to find new sources of CP-violation.

\subsubsection{Permanent Electric Dipole Moments (EDMs)}

A permanent electric dipole moment of any fundamental particle
violates both parity and time reversal (T) symmetries. 
With the assumption of 
CPT invariance a  permanent dipole moment also violates CP.
Permanent electric dipole moments  for all particles are caused by CP violation as it is known from 
the K systems through higher order loops. These are at least 4 orders of magnitude below the
present experimentally established limits. Indeed, a large number of speculative models
foresees permanent electric dipole moments which could be as large as 
the present experimental limits just allow. 
Historically the non-observation of permanent electric dipole moments has
ruled out more speculative models than any other experimental
approach in all of particle physics \cite{Ramsey_1999}. 
\begin{table}[ht]
\caption{\it The best limits on permanent electric dipole moments.} 
{\footnotesize
\begin{tabular}{|c|c|c|} \hline

 Particle & Limit/Measurement (e-cm) & Method \\   \hline
e               & $<1.6 \times 10^{-27}$  &  Thallium beam \cite{Regan} \\
$\mu$           & $<2.8\times 10^{-19}$ &  Tilt of precession plane in magnetic moment experiment \cite{BNL_EDM} \\ 
$\tau$          & $(-2.2 <d_{\tau}<4.5)\times 10^{-17}$ &  BELLE  $e^{+} e^{-} \rightarrow \tau \tau $ events \cite{Belle_tau} \\ 
n               & $<6.3 \times 10^{-26}$ & Ultra-cold neutrons \cite{Harris}  \\ 
p               & $(-3.7\pm 6.3) \times 10^{-23}$ & 120kHz thallium spin resonance  \cite{Proton_EDM}  \\ 
$\Lambda$       & $(-3.0\pm 7.4) \times 10^{-17}$ &  Tilt of precession plane in magnetic moment experiment \cite{Lambda_EDM} \\ 
$\nu_{e,\mu}$   & $<2 \times 10^{-21}$ &  Inferred from magnetic moment limits \cite{delAguila} \\ 
$\nu_{\tau}$    & $<5.2 \times 10^{-17}$ & Z decay width \cite{NuTau_EDM}  \\   
Hg-atom         & $< 2.1 \times 10^{-28}$ & mercury atom spin precession \cite{Romalis_2001}\\ \hline

\end{tabular}
}
\label{EDMs} 
\end{table}

Permanent electric dipole moments have been
searched for in various systems with different sensitivities 
(Table \ref{EDMs}). In composed systems such as molecules
or atoms fundamental particle dipole moments of constituents may be
significantly enhanced\cite{Sandars_2001}. Particularly in polarizable
systems there can exist large internal fields.

There is no preferred system to search for an EDM. In fact,
many systems need to be examined, because depending
on the underlying process different systems have
in general quite significantly different susceptibility
to acquire an EDM through a particular mechanism.
In fact, one needs to investigate different systems.
An EDM may be found an ''intrinsic property'' of
an elementary particle asw we know them, because the underlying 
mechanism is not accessible at present. However, it can also
arise from CP-odd forces between the constituents under observation,
e.g. between nucleons in nuclei or between nuclei and
electrons. Such EDMs could be much higher than such
expected for elementary particles originating within the popular, usually 
considered standard theory models. No other constraints are known.

\begin{figure}[hbt]
   \centering
   \epsfig{figure=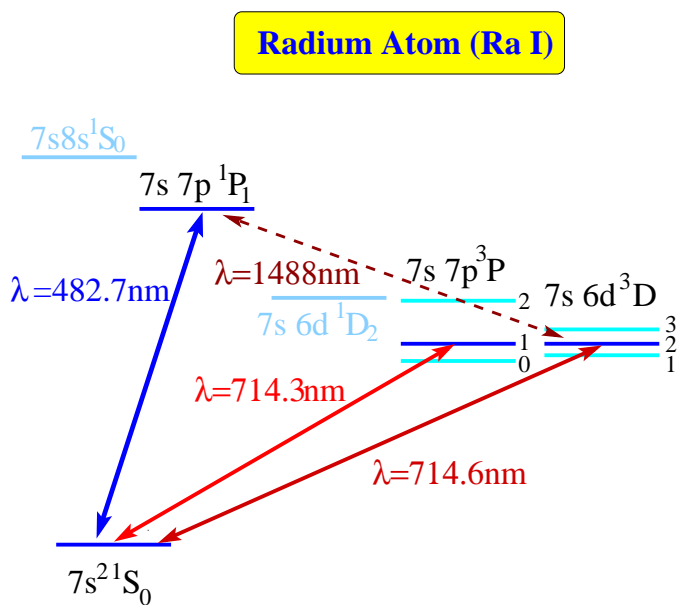,angle=0,width=3.5in}
   \caption{\it The Ra atom has two levels of opposite parity
   in the triplett system, which can result in a strong enhancement
   factor for a possible electron EDM
   \cite{Dzuba_2001}.
   }
\label{radium}
\end{figure}

In this active field of research we had recently a number of novel 
developments.
One of them concerns the Ra atom, which  
has rather close lying $7s7p^3P_1$ and $7s6d^3D_2$ states (Fig. \ref{radium}).
Because they are of opposite parity, a significant enhancement 
has been predicted  for an  electron EDM \cite{Dzuba_2001}, much
higher than for any other atomic system. Further more,
many Ra isotopes are in a region where (dynamic) octupole deformation occurs
for the nuclei, which also may enhance the effect of a nucleon EDM
substantially, i.e. by some two orders of magnitude.
From a technical point of view the Ra atomic levels of interest for en experiment
are well accessible spectroscopically and a variety of isotopes can be produced 
in nuclear reactions. 
The advantage of an accelerator based Ra experiment is apparent,
because EDMs require isotopes with spin
and all Ra isotopes with finite nuclear spin are relatively short-lived
\cite{Jungmann_2002}.

\begin{figure}[bt] 
\begin{minipage}{3in} 
   \centering
   \epsfig{figure=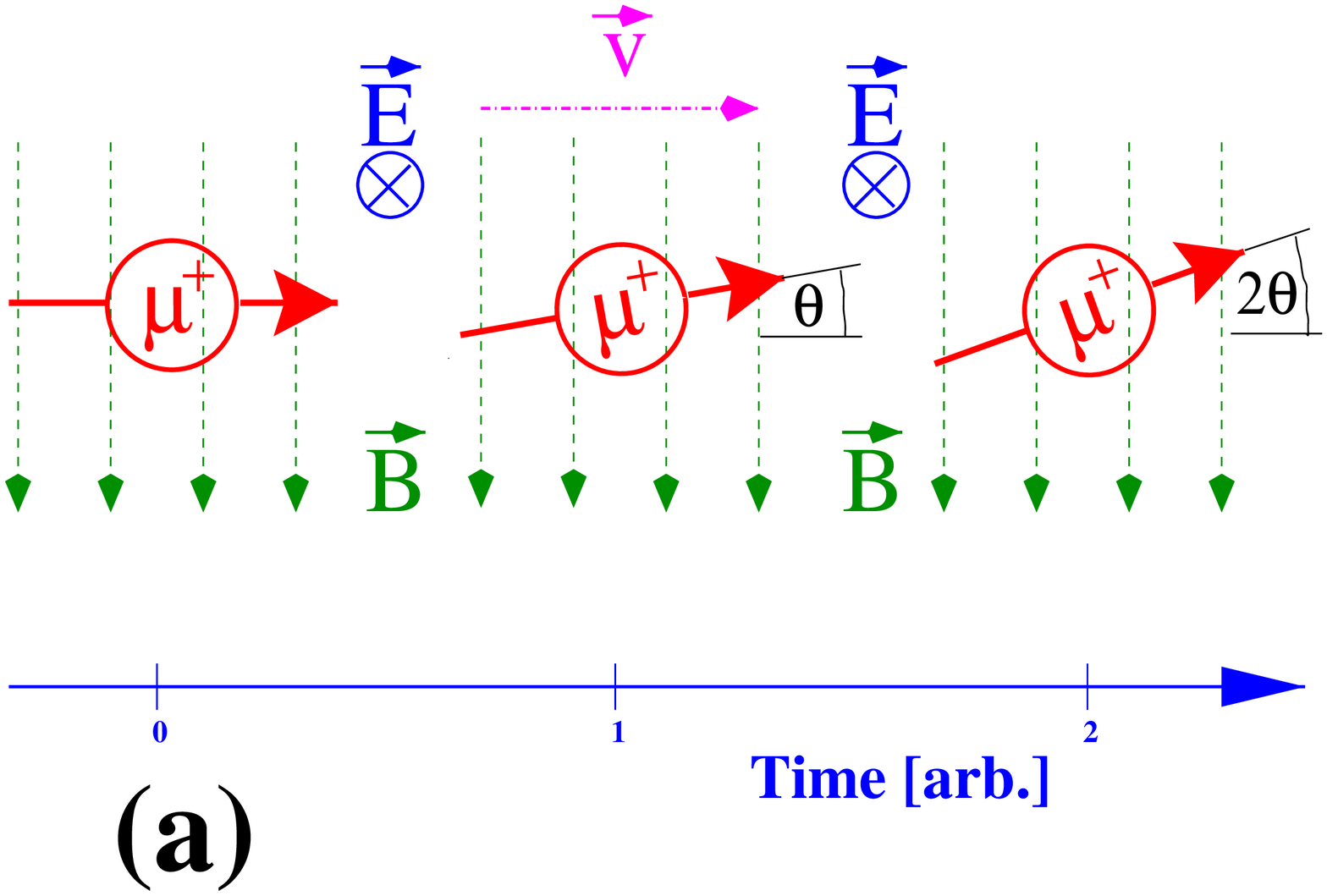,angle=0,width=\textwidth}
\end{minipage}
\hfill 
\begin{minipage}{3in}
   \centering
   \epsfig{figure=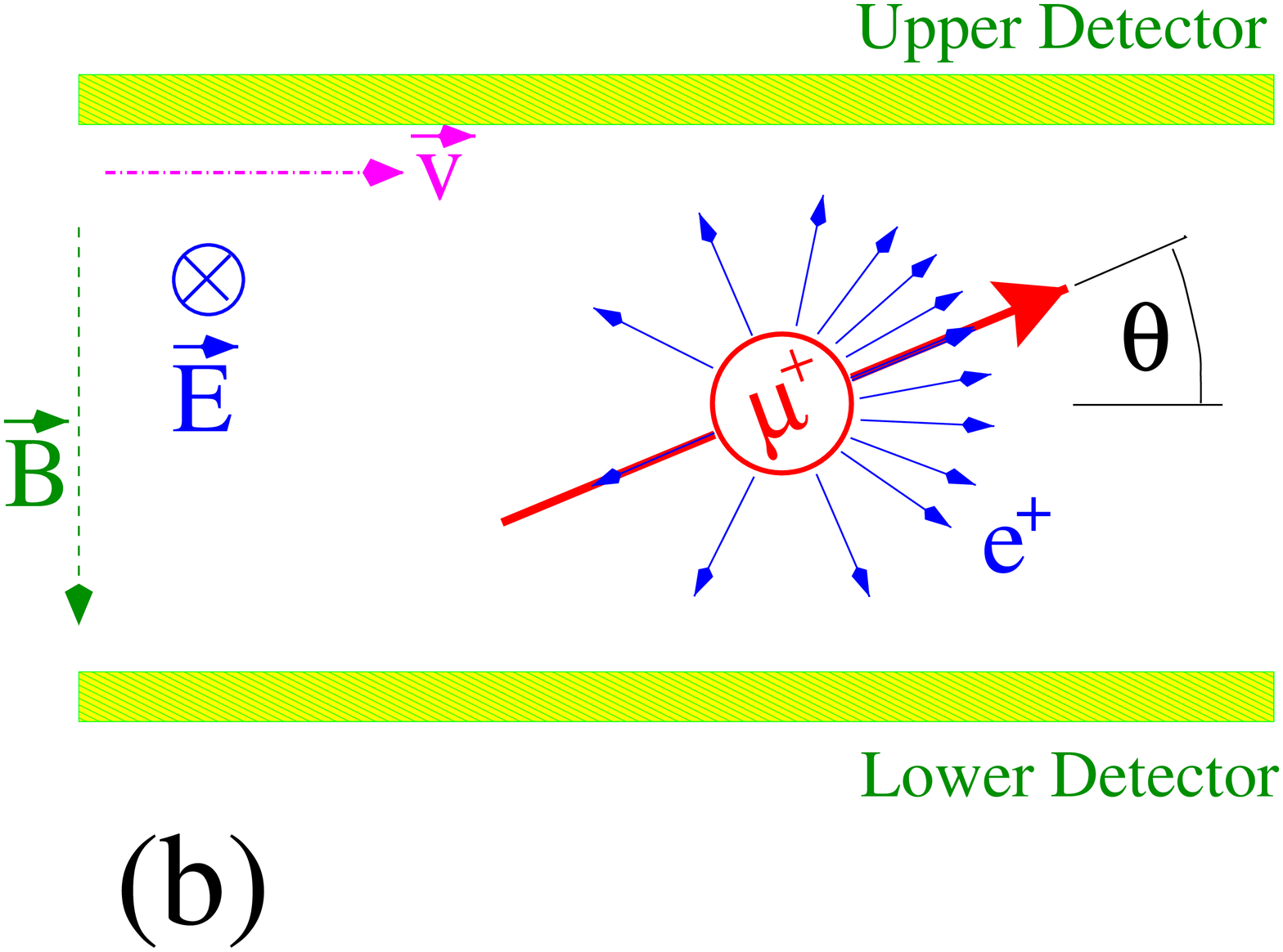,angle=0,width=\textwidth}

\end{minipage}
   \caption{
   \it
     (a) The principle of a ring EDM search experiment. The spin of
     a fast moving charged particle with an EDM rotates
     around an axis along the radius of a magnetic storage ring
     due to the motional electric field.
     (b) For muons the spin rotation translates into a time dependent
    change of the count rate observed by counters for the 
    positrons from muon decays which are located above and below 
    the plane of orbit \cite{Farley_2004}.
      }
\label{edm_principle}
\end{figure}

A very novel idea was introduced for measuring an 
EDM of charged particles. In this method the high
motional electric field is exploited, which charged particles at relativistic speeds 
experience in a magnetic storage ring.
(Fig. \ref{edm_principle} a). In such an experiment the Schiff theorem can be circumvented
(which had excluded charged particles from experiments because of the
Lorentz force acceleration) because of the non-trivial geometry of the problem
\cite{Sandars_2001}. With an
additional radial electric field in the storage region the spin precession due to the
magnetic moment anomaly can be compensated, if the 
effective magnetic anomaly $a_{eff}$ is small, i.e. $ a_{eff}<<1$. 
The method was first considered for muons. For longitudinally polarized 
muons injected into the ring an EDM
would express itself 
as a  spin rotation out of the orbital plane (Fig. \ref{edm_principle} b).
This can be observed as a time dependent (to first oder linear in time) 
change of the above/below  the plane of orbit counting rate ratio. 
For the possible muon beams at the future J-PARC facility in Japan
a sensitivity of $10^{-24}$e\,cm is expected \cite{Yannis_2003}. 
In such an experiment the possible muon flux is a major limitation.
For models with nonlinear mass scaling of EDM's such an experiment would 
already be more sensitive to certain new physics models
than the present limit on the electron EDM 
\cite{Babu_2000}. 
An experiment carried out at a more intense muon source could provide
a significantly more sensitive probe to CP violation in the second 
generation of particles without strangeness.

The deuteron is the simplest known nucleus. Here an EDM
could arise not only from a proton or a neutron EDM, but also
from CP-odd nuclear forces. It was shown very recently \cite{Liu_2004} that
the deuteron can be in certain scenarios significantly more sensitive than the
neutron. In equation (\ref{nedm}) this situation is
evident for the case of quark chromo-EDMs:
\begin{eqnarray}
d_{\mathcal{D}} & = & -4.67\, d_{d}^{c}+5.22\, d_{u}^{c}\,,\nonumber \\
d_{n} & = & -0.01\, d_{d}^{c}+0.49\, d_{u}^{c}\,.\label{nedm}
\end{eqnarray}
It should be noted that because of its rather small magnetic anomaly
the deuteron is a particularly interesting candidate for a ring EDM experiment
and a proposal with a sensitivity of $10^{-27}$~e\,cm exists \cite{Semertzidis_2004}.
In this case scattering off a target will be used to observe a spin precession.
As possible sites of an experiment the Brookhaven National Laboratory (USA),
the Indiana University Cyclotron Facility (USA) and the Kernfysisch Versneller
Instituut (Netherlands) are considered.

\subsubsection{Correlations in $\beta$-decays}

In standard theory the structure of weak
interactions is V-A, which means there are vector (V) and axial-vector (A) 
currents with opposite relative sign causing a left handed structure 
of the interaction and parity violation \cite{Herczeg_2001}. 
Other possibilities like scalar, pseudo-scalar and tensor type 
interactions which might be possible would be clear 
signatures of new physics. So far they have been searched 
for without positive result. However, the bounds on parameters
are not very tight and leave room for various speculative possibilities. 
The double differential decay probability 
$ d^2W/d\Omega_e d\Omega_{\nu}$for a $\beta$-radioactive nucleus is
related to the electron and neutrino momenta $\vec{p}$ and $\vec{q}$ through
\begin{eqnarray}
\label{diffprob}
\frac{d^2W}{d\Omega_e d\Omega_{\nu}} & \sim & 1 +  a ~\frac{\vec{p}\cdot\vec{q}}{E} 
+  b ~~\sqrt{1-(Z \alpha)^2}~~\frac{m_e}{E}       \nonumber     \\
& & + <\vec{J}>     \cdot \left[ A~~ \frac{\vec{p}}{E} + B~~\vec{q} + D~~\frac{\vec{p} \times ~\vec{q}}{E} \right]\\
& &+ <\vec{\sigma}> \cdot \left[ G~~ \frac{\vec{p}}{E} + Q~~\vec{J} + R~~ <\vec{J}> \times ~\frac{\vec{q}}{E} \right] \nonumber
\end{eqnarray}
where   $m_e$ is the $\beta$-particle mass,
        $E$ its energy,
        $\vec{\sigma}$ its spin,  and
        $\vec{J}$ is the spin of the decaying nucleus.
        The coefficients D and R are studied in a number of
        experiments at this time and they are T violating in nature. 
        Here D is of particular interest for
further restricting model parameters. It describes the correlation between 
the neutrino and $\beta$-particle momentum vectors for spin polarized nuclei. 
The coefficient R is highly sensitive within a smaller set of 
speculative models, since in this region there exist some already well established
constraints, e.g., from searches for permanent electric dipole moments  \cite{Herczeg_2001}. 

From the experimental point of view, 
an efficient direct measurement of the neutrino momentum is not possible.
The recoiling nucleus can be detected instead and the neutrino 
momentum can be reconstructed using the  kinematics of the process.
Since the recoil nuclei have typical energies in the few 10 eV range,
precise measurements can only be performed, if the decaying isotopes are 
suspended using extreme shallow potential wells. Such exist, for example,
in magneto-optical traps, where many atomic species can be stored at 
temperatures below 1 mK. 

Such research is being performed at a number of laboratories worldwide.
At the Kernfysisch Versneller Instituut (KVI) in Groningen a new facility
is being set up, in which T-violation research will be a central scientific issue
\cite{Jungmann_2002,Turkstra_2002}. At this new facility
the isotopes of primary interest are $^{20,21}$Na and $^{18,19}$Ne.
These atoms have suitable spectral lines for 
optical trapping and since also the nuclear properties are such that
rather clean transitions can be observed.

A recent measurement at Berkeley, USA, the asymmetry parameter $a$ in the 
$\beta$-decay of  $^{21}$Na has been measured in optically trapped atoms
\cite{Scielzo_2004}. The value
differs from the present SM value by about 3 standard deviations.
Whether this is an indication of new physics 
reflected in new interactions in $\beta$-decay, this depends strongly
on the $\beta / (\beta +\gamma)$ decay branching ratio  for which
some 5 measurements exists which in part disagree significantly \cite{Endt_1990} 
New measurements are needed.

\section{Properties of Known Basic Interactions}

\subsection{Electromagnetism and Fundamental Constants}

In the electro-weak part of the SM
very high  precision can be achieved for calculations,
in particular within Quantum Electrodynamics (QED), which is
the best tested field theory we know and a key element of the SM.
QED allows for extracting  accurate values of important fundamental 
constants from high precision experiments on free particles 
and light bound systems, where perturbative approaches work
very well for their theoretical description. Examples are the
fine structure constant $\alpha$ or the Rydberg constant R$_{\infty}$.
The obtained numbers are needed to describe the known 
interactions precisely. 
Furthermore, accurate calculations 
provide a basis to searches for deviations from SM predictions.
Such differences would reveal clear and undisputed signs of New Physics
and hints for the validity of speculative extensions to the SM.
For bound systems containing nuclei with high electric charges 
QED resembles a field theory with strong coupling and new 
theoretical methods are needed.

\subsubsection{Muonium and Muon Magnetic Anomaly}

The interpretation of measurements in the 
muonium atom, the bound
state of a $\mu^+$ and an $e^-$, is free of difficulties
arising from the structure of its constituents \cite{Jungmann_2004}. 
Thus QED predictions with two orders of 
magnitude higher accuracy than for the hydrogen atom are 
possible.
The ground state hyperfine splitting as 
well as the $1s-2s$ energy difference have been precisely determined 
recently. These measurements 
can be interpreted as QED tests or alternatively  
-assuming the validity of QED- 
as independent  measurements  of $\alpha$ as well as of muon properties 
(muon mass $m_{\mu}$ and muon magnetic moment $\mu_{\mu}$).
These experiments are statistics limited. Significantly
improved values would be possible at new intense muon sources.
There is a close connection between muonium spectroscopy
and a measurement of the muon magnetic anomaly $a_{\mu}$,
the relative deviation of the muon g-factor from the Dirac value 2.
Muonium spectroscopy provides fundamental muon constants, such as its mass,
electric charge and magnetic moment.

\begin{figure}[hbt]
   \centering
   \epsfig{figure=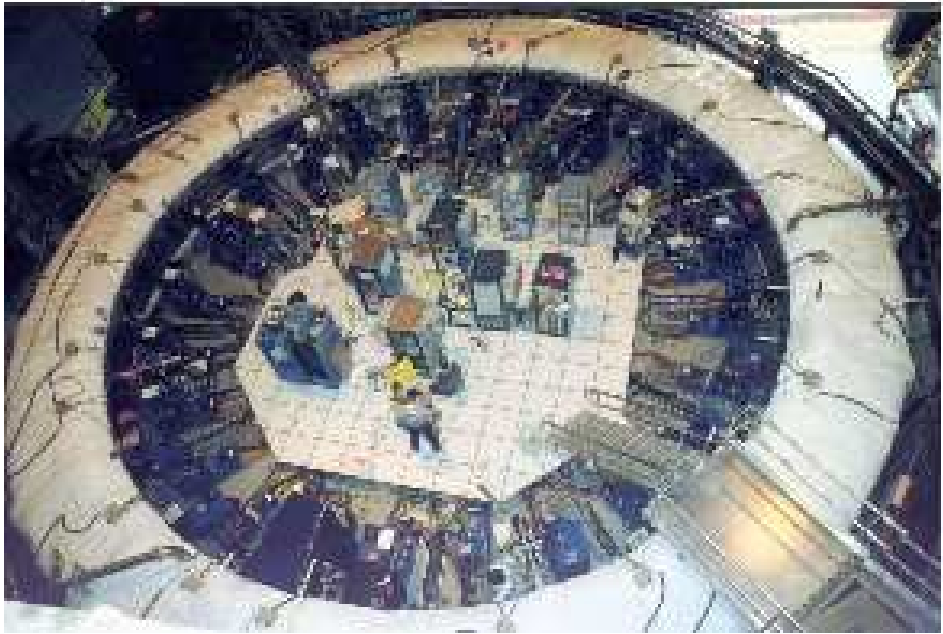,angle=0,width=5in}
   \caption{\it The g-2 storage ring at the Brookhaven National Laboratory
   \cite{Bennett_2004}. 
   }
\label{ring}
\end{figure}

Precise values of these fundamental constants are indispensable 
for the evaluation of the experimental results of
a muon g-2 measurement series 
in a magnetic storage ring (Fig. \ref{ring})
at the Brookhaven National Laboratory \cite{Bennett_2004}.
The muon magnetic anomaly
arises from quantum effects and is mostly due to QED.
Further, there is a contribution from strong interactions of 58~ppm 
which arises from hadronic vacuum polarization. 
The influence of weak interactions amounts to 1.3 ppm. 
Whereas QED and weak effects can be calculated from first principles,
the hadronic contribution
needs to be evaluated through a dispersion relation
and experimental input from $e^+$-$e^-$ annihilation into hardrons
(up to now in the essential region from the CMD experiment in Novosibirsk, Russia)
 or hadronic 
$\tau$-decays.  Calculations of the hadronic
part in $a_{\mu}$ depend on the choice of presently available 
experimental hadronic data. The results for $a_{\mu}$
differ by 3.0
respectively 1.6 standard deviations from the averaged experimental value.
Intense theoretical and experimental efforts are needed to solve 
the hadronic correction puzzle. The available new data 
on $e^+$-$e^-$ annihilation from the KLOE experiment in Frascati, Italy, appear to confirm earlier
measurements \cite{Aloisio_2004}.
For the muon magnetic anomaly improvements both in theory and
experiment are required, before a definite conclusion can be drawn
whether a hint of physics beyond standard theory \cite{Chavez_2004}  
has been seen. A continuation of the g-2 experiment with improved 
equipment and beams
was scientifically approved in 2004.

\subsubsection{Muonic Hydrogen and Proton Radius}

A measurement of the proton mean square charge radius ${\rm r}_p$ is underway
at PSI \cite{Kottmann_2002}. The experiment aims for a determination
of the classical Lambshift in the n=2 state with laser spectroscopy.
The transition is within reach of infrared laser radiation.
Muonic hydrogen has a higher sensitivity to
proton properties compared to natural  hydrogen 
owing to the about 200 times smaller Bohr radius of the system and
the associated higher overlap probability of the muon with
the proton (as compared to the electron in natural hydrogen) .
One expects a significant improvement in the knowledge of ${\rm r}_p$
over the value available from high precision laser spectroscopy 
in the 1s-2s transition in natural hydrogen.

\subsubsection{Does $\alpha_{QED}$ Vary with Time ?}

The question whether fundamental constants are stable in time 
goes back to the large number hypothesis of Dirac. 
More recently reports came out, in which evidence for a time variation
of the fine structure constant $ \alpha_{QED}$  was reported from astronomical
observations. Absorption of quasar light in interstellar media was employed
to search for shifts in atomic fine structure lines. Due to relativistic effects,
some atomic levels would shift positive and some others would simultaneously shift negative,
if $\alpha_{QED}$ would vary \cite{Angstmann_2004}. A variation
of ${\dot{\alpha}/\alpha} = (6.40 \pm 1.35) \times 10^{-16}{\rm \,yr}^{-1}$
is observed by one collaboration \cite{Murphy_2003}. However, this could not be
confirmed by a second group working with the same astronomical instruments 
\cite{Murphy_2004} and by laboratory experiments using 
the narrow atomic hydrogen 1s-2s transition in a comparison with
an atomic clock \cite{Fischer_2004}. Further work to clarify the situation in this 
lively subfield will be needed.

\subsection{Quantum Chromodynamics (QCD)}

At PSI spectroscopy of atomic x-rays from pionic hydrogen allows to determine
a Strong Interaction shift and broadening of atomic transitions. For the chosen
3p-1s transition the shift and broadening are predominantly due to the 1s state. 
The measurement  yields the pion-nucleon scattering length with very good precision
\cite{Gotta_2003} and therefore presents a  high precision test of chiral perturbation theory,
a powerful low energy approach in QCD.

 \begin{figure}[hbt]
   \centering
   \epsfig{figure=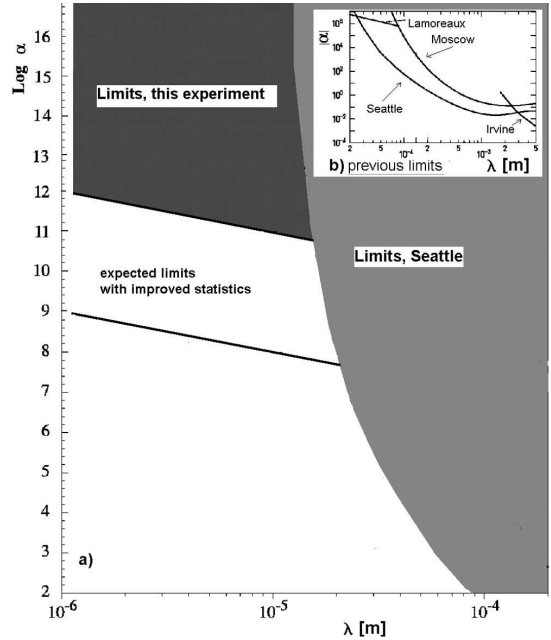,angle=0,width=3in}
   \caption{\it Limits on non-Newtonian gravity. The strength $|\alpha|$
   as a function of the 
   the Yukawa length scale. The parameter space was further
   limited by a neutron gravitation interferometer \cite{Abele_2003}.
   }
   \label{grav}
\end{figure}

\subsection{Gravity}

String and M theories try to find a common description of
gravity and quantum mechanics. In their context appear predictions
of extra dimensions which could manifest 
\begin{sidewaystable}[t]\center 
 \caption
 {\it Research areas and typical experiments where  
  a high power proton driver is indicated for 
  significant progress \cite{CERN_PD_2004}.
 The necessary typical proton beam energy is given. A
 beam power of about 4 MW is assumed. Most experiments would benefit from a 
 pulsed time structure of the beam.}
{ \footnotesize
\begin{tabular}{|l|c|c|c|c|c|}
 \hline
\hspace*{0mm} Research Field & Physics Question addressed& Method & Comments & $\approx$ 1~GeV & $\approx$ 30~GeV\\
\hline  \hline
nature of & oscillations, CP violation& long baseline & novel detectors? Salt domes ?  & $\times$  & $\times $\\
neutrinos & masses                    & spectrometer  &   only $\nu_{\mu}$             &     --    & $\times $\\
\hline  
T and CP  & permanent electric& spin precession in    & novel method using storage rings& $\times$ & $\times$ \\
violation & dipole moments;   & electric fields;      & Radium atoms;                   & & \\
          & D (R) coefficients& trapping of radioactive& stored radioactive atoms;      & $\times$  & $\times$ \\
          & in $\beta$-decays;& atoms;                &   &   &  \\
          & D$^0$-decays      & spectrometer          & antiproton facility & --  & $\times$ \\
\hline          
rare and  & n-$\overline{\rm n} $ conversion & dedicated     & ultracold n's  &  $\times$ &  \\
forbidden & M-$\overline{\rm M} $ conversion & spectrometers & novel method, unique potential  & $\times$  &  \\
decays    & $\mu \rightarrow e \gamma$     &               & unique potential  & $\times$  &  \\ 
          & $\mu \rightarrow 3e $          &               & unique potential  & $\times$  &  \\
          & $\mu {\rm N} \rightarrow e {\rm N}$ &          & unique potential  & $\times$  &  \\
\hline
Correlations     & non V-A in    & radioactive    & optically trapped   & $\times$ & $\times$   \\ 
in $\beta$-decays& $\beta$-decay & nuclear decays & radioactive isotopes& $\times$ & $\times$   \\
\hline
unitarity of &  n-decay           & lifetimes                & large potential       & $\times$ & \\
CKM          &  $pi$-$\beta$-decay&   and                    & to test SM            & $\times$ & \\
matrix       &  K-decays          & transition probabilities & in new precision round&  --      & $\times $\\
\hline
CPT          &   nuclei                & siderial variations               & interaction based  & $\times$ & \\
conservation &   p, $\overline{\rm p}$ & of spin dependent quantities;     & models needed      &   --     &$\times $ \\
             &   $\mu$                 & particle-antiparticle properties&                    & $\times$ &$\times $ \\
             \hline 
\end{tabular}
}
\label{proton_driver}  
\end{sidewaystable} 
\clearpage
\noindent
themselves in deviations
from the Newtonian laws of gravity at small distances. Therefore 
an number of searches for such large extra dimensions has been
started. At the Institute Laue langevain in Grenoble, France,
a new limit in parameter space (Fig. \ref{grav}) has been established  for 
extra forces of the type 
\begin{equation}
V(r)= G \frac{m_1 \cdot m_2}{r} (1+ \alpha \cdot \exp(\frac{-r}{\lambda}) ,
\end{equation}
where $\alpha$ determines the strength and $\lambda$ is the Yukawa range
of the additional interaction.
The experiment uses
quantum mechanical interference patterns from
ultra-cold neutrons which may be viewed as 
gravitational matter "standing" waves \cite{Nesvizhevsky_2003}.

\section{New Instrumentation Needed}

Progress in the field of low energy experiments to verify and test the SM
and to search for extensions to it would benefit in many cases 
significantly from new instrumentation and a new generation of particle sources. 
In particular, a high power proton driver would boost a large number of 
possible experiments which all have  a high and robust  discovery potential \cite{NUPECC_2004}.
In Table \ref{proton_driver} two possible scenarios for a 1~GeV
and a 30~GeV machine are compared.
The availability of such a machine would be desirable for a number 
of other fields as well, such as neutron scattering, in particular ultra-cold neutron research,
or a new ISOL facility (e.g. EURISOL)
for nuclear physics with nuclei far off the valley of stability. A joint
effort of several communities 
could benefit from synergy effects. Possibilities for such a machine 
could arise at CERN \cite{CERN_PD_2004,Aysto_2001}, FEMILAB, J-PARC and GSI
with either a high power linac or a true rapid cycling synchrotron.

\section{Conclusions}

 Nuclear physics and nuclear techniques offer a variety of possibilities to investigate fundamental 
 symmetries in physics and to search for physics beyond the SM. 
 Experiments at Nuclear Physics facilities at low and 
intermediate energies 
offer in this 
respect a variety of possibilities which are complementary 
to approaches in High Energy physics and in some cases
exceed those significantly in their potential to steer
physical model building.

 The advantage of high particle fluxes at a Multi-Megawatt facility allow
 higher sensitivity to rare processes because of higher statistics 
 and because also in part novel experimental approaches are enabled by the combination of
 particle number and an appropriate time structure of the beam. The field is looking forward
 to a rich future.

\section{Acknowledgments} 

The author would like to the members of the  NuPECC
Long Range Plan 2004 Fundamental Interaction working group \cite{NUPECC_2004}
for numerous fruitful discussions.
This work was supported in part by the Dutch Stichting
voor Fundamenteel Onderzoek der Materie (FOM) in the
framework of the TRI$\mu$P programme.

\end{document}